\documentclass[useAMS,usenatbib]{mnras}

\usepackage[utf8]{inputenc}
\usepackage{graphicx}
\usepackage{color}
\usepackage{booktabs}
\usepackage[nolist]{acronym}
\usepackage{amssymb}
\usepackage[]{natbib}
\usepackage{float}
\usepackage{hyperref}
\usepackage{amsmath,verbatim,mathtools,needspace,enumitem,etoolbox,physics,microtype}
\usepackage{epsfig}
\usepackage{url}
\usepackage{multirow}
\usepackage{array,multirow}

\newcolumntype{M}[1]{>{\centering\arraybackslash}m{#1}}
\newcolumntype{N}{@{}m{0pt}@{}}

\newcommand{\eagle}{{\sc eagle}}
\newcommand{\mobse}{{\sc mobse}}

\def\apj{ApJ}

\def\apjl{ApJ}

\def\mnras{MNRAS}

\def\nat{Nature}
\def\araa{ARA\&A}
\def\aap{A\&A}

\def\pc{{\rm\thinspace pc}}
\def\kpc{{\rm\thinspace kpc}}

\def\Mpc{{\rm\thinspace Mpc}}

\def\Msun{\hbox{$\rm\thinspace M_{\odot}$}}

\def\yr{{\rm\thinspace yr}}

\def\Msunpc2{{\Msun\pc}^{-2}}
\def\Msunyrkpc2{{\Msun\yr^{-1}\kpc}^{-2}}

\def\magarcsec2{{\rm\thinspace mag\thinspace arcsec}^{-2}}

\begin{document}

\title[An astrophysically motivated ranking criterion for EM follow-up of GW events]{An astrophysically motivated ranking criterion for low-latency electromagnetic follow-up of gravitational wave events}

\author[Artale et al.]{M. Celeste Artale$^{1}$\thanks{E-mail:Maria.Artale@uibk.ac.at, mcartale@gmail.com}, Yann Bouffanais$^{2,3}$\thanks{E-mail:yann.bouffanais@gmail.com}, Michela Mapelli$^{2,3,4}$,
Nicola Giacobbo$^{2,3,4}$, \newauthor   Nadeen B. Sabha$^{1}$,  Filippo Santoliquido$^{2,3}$, Mario Pasquato$^{3,4}$, \& Mario Spera$^{2,3,4,5,6}$\\ 
$^{1}$Institut f\"{u}r Astro- und Teilchenphysik, Universit\"{a}t Innsbruck, Technikerstrasse 25/8, 6020 Innsbruck, Austria\\
$^{2}$Physics and Astronomy Department Galileo Galilei, University of Padova, Vicolo dell'Osservatorio 3, I--35122, Padova, Italy\\
$^{3}$INFN-Padova, Via Marzolo 8, I--35131 Padova, Italy\\
$^{4}$INAF-Osservatorio Astronomico di Padova, Vicolo dell'Osservatorio 5, I--35122, Padova, Italy\\
$^{5}$Center for Interdisciplinary Exploration and Research in Astrophysics (CIERA), Evanston, IL 60208, USA \\
$^{6}$Department of Physics \& Astronomy, Northwestern University, Evanston, IL 60208, USA \\
}
\maketitle
\begin{abstract}
 We investigate the properties of the host galaxies of  compact binary mergers across cosmic time. 
  To this end, we combine population synthesis simulations together with galaxy catalogues from the hydrodynamical cosmological simulation \eagle\ to derive the properties of the host galaxies of binary neutron star (BNS), black hole-neutron star (BHNS) and binary black hole (BBH) mergers. Within this framework, we derive the host galaxy probability, i.e., the probability that a galaxy hosts a compact binary coalescence as a function of its stellar mass, star formation rate, $K_s$ magnitude and $B$ magnitude. This quantity is particularly important for low-latency searches of gravitational wave (GW) sources as it provides a way to rank galaxies lying inside the credible region in the sky of a given GW detection, hence reducing the number of viable host candidates. Furthermore, even if no electromagnetic counterpart is detected, the proposed ranking criterion can still be used  to classify the galaxies contained in the error box.
  Our results show that massive galaxies (or equivalently galaxies with a high luminosity in $K_s$ band) have a higher probability of hosting BNS, BHNS, and BBH mergers.  We provide the probabilities in a suitable format to be implemented in future low-latency searches.

\end{abstract}


\section{Introduction}
The number of gravitational wave (GW) detections is rapidly growing. The LIGO-Virgo collaboration (LVC) reported eleven compact binary mergers from their first two observing runs: ten merging binary black holes (BBHs) and one binary neutron star (BNS) \citep{Abbott2016a,Abbott2016b,Abbott2016c,Abbott2017,Abbott2017b,abbottGW170814,AbbottO2,AbbottO2popandrate}. Based on the results of an independent pipeline, \cite{venumadhav2019} and \cite{zackay2019} claimed several additional GW candidates. GW190425, the second BNS \citep{AbbottGW190425}, and GW190412, the first unequal-mass BBH \citep{AbbottGW190412}, have already been reported as first results of the third observing run (O3).  By the end of O3, we expect to have several tens of BBHs, several new BNSs, and potentially the first merging black hole -- neutron star binary (BHNS). 

\vspace{1cm}
The electromagnetic (EM) counterpart of the first detected BNS merger, GW170817 \citep{Abbott2017b}, was observed through the entire EM spectrum, from radio to gamma-ray wavelengths \citep{abbottmultimessenger,abbottGRB,goldstein2017,savchenko2017,margutti2017,coulter2017,soares-santos2017,chornock2017,cowperthwaite2017,nicholl2017,pian2017,alexander2017}. This definitely confirmed the association between BNS mergers, short gamma ray bursts and kilonovae.
Moreover, with the improvement of GW interferometers, more multimessenger detections are expected in the near future. 
Nonetheless, finding the EM counterpart is a "needle in the haystack'' problem, because the uncertainty on the sky localization for current GW detector network spans from tens to thousands 
of square degrees \citep{Abbott2018}. 

The standard method for EM follow-up, uses the probability skymap error box reported by LIGO-Virgo for each GW source \citep{Veitch2015,Singer2016}.
In order to improve the search process, one strategy is to first identify the galaxies contained within the credible regions given by the LVC using a complete galaxy catalogue such as GLADE \citep[][]{Dalya2018}, and then rank the galaxies according to their localisation probability \citep[see e.g.,][]{Gehrels2016,Arcavi2017}. 
Finding a criterion to predict the properties of the most likely host galaxies of a GW event might improve the overall ranking procedure in two ways. On the one hand, it can speed up low-latency searches for EM counterparts by prioritising observations of the most likely host galaxies. On the other hand, even if the counterpart is not detected, it can be used to weigh the candidate galaxies identified in the error box, facilitating some further analysis such as the measurement of the Hubble constant \citep{abbottH02019}. 
Previous works propose to rank the most likely host galaxies according to their properties such as luminosity and distance \citep[see e.g.,][]{Nuttall2010,Fan2014,Antolini2016}.
In particular,  \citet{Ducoin2019}  have recently shown that including a dependence on  galaxy stellar mass in the probability density improves the galaxy ranking dramatically.

From a theoretical perspective, predicting the most likely host galaxies of compact binary mergers is a challenging task, due to the extreme physical scales involved.
A possible approach to tackle this problem is to combine galaxy formation models with binary population synthesis simulations. This methodology has been implemented in several works using semi-analytic models or cosmological simulations together with population synthesis models 
\citep[see e.g.,][]{OShaughnessy2010,OShaughnessy2017,Dvorkin2016,Lamberts2016,Mapelli2017,Schneider2017,
Cao2018,Mapelli2018,Mapelli2018b,Artale2019,Artale2019b,
Marassi2019,Boco2019,Conselice2019,Eldridge2019,Mapelli2019,Toffano2019,Adhikari2020,Jiang2020}.

In our previous studies \citep{Artale2019,Artale2019b}, 
we found a strong correlation between the binary merger rate per galaxy, the stellar mass and the star formation rate (SFR) of the host galaxy.
In particular, we showed that the correlation between merger rate per galaxy and stellar mass is the strongest one, is almost linear, holds for all different types of compact binaries (BBHs, BHNSs and BNSs) and for the entire redshift range considered, from $z=0$ to $z=6$.
Moreover, the median stellar mass of host galaxies increases as redshift decreases, as a consequence of the cosmic assembly history of the galaxies, together with the delay time of merging compact binaries\footnote{We refer to delay time as the time elapsed between the formation of the stellar progenitor to the merger of the binary compact object system.}.

In this study, we make use of the simulations already presented in \cite{Artale2019} and \cite{Artale2019b} to compute the density of mergers per galaxy as a function of its stellar mass and SFR at different redshifts (up to redshift $z\sim{}6$). From there, we use the density to build a probability-like quantity (host galaxy probability) that can be integrated as an additional ranking criterion in low-latency searches of EM counterparts, facilitating a prompt multi-messenger follow-up 
\citep{nissanke2013,hanna2014,Gehrels2016,Mapelli2018b,delpozzo2018}. Alternatively, if the EM counterpart is not detected, our results can still be used to rank possible host galaxies in the LVC error box, hence providing useful information for measurements of the Hubble constant (e.g. \citealt{Schutz1986,delpozzo2012,LIGO2017StandardSirens,Chen2018,Vitale2018,Fishbach2019,Soares-Santos2019,delpozzo2018,abbottH02019}).

 This paper is organized as follows. Section~\ref{sec:method} describes the  methodology we implemented to seed the \eagle\ cosmological simulation with a population of merging compact objects (BNSs, BHNSs, and BBHs), along with the approach we used to build our ranking criterion. We present our results in Section~\ref{sec:results} and we compare them with previous work in Section~\ref{sec:discussion}. Our main conclusions are  drawn in Section~\ref{sec:conclusions}.

\section{Method}\label{sec:method}

\subsection{Catalogues of binary mergers and their host galaxies}
We adopt the methodology introduced by \citet{Mapelli2017} \citep[also implemented in][]{Mapelli2018,Mapelli2018b,Mapelli2019,Artale2019,Artale2019b}
to produce a galaxy catalogue of merging compact objects. 
The main idea is to combine the simulated galaxies from the \eagle\ simulation suite \citep{Schaye15} with merging compact object catalogues obtained with the population synthesis code {\sc mobse} \citep{Giacobbo2018}. In this section we summarize the main details of the methodology; while a more exhaustive description can be found in the aforementioned papers.

The population synthesis code {\sc mobse} \citep{Giacobbo2018} is an upgrade of the {\sc bse} code \citep{Hurley2000,Hurley2002}.
{\sc mobse} includes new prescriptions for metallicity dependent stellar winds \citep[][]{Vink2001,Vink2005,Chen2015}, core collapse supernovae (SNe, following \citealt{Fryer2012}), and pair-instability and pulsational pair-instability SNe \citep{Woosley2017,Spera2017}.
In this work, we use the catalogues from the merging compact model CC15$\alpha{}$5 presented in \citet{Giacobbo2018B}, which was already implemented in several previous works \citep{Mapelli2018,Mapelli2019,Artale2019,Toffano2019,Artale2019b}.
The run CC15$\alpha{}$5 is composed of 12 subsets of metallicities $Z = 0.0002$, 0.0004, 0.0008, 0.0012, 0.0016, 0.002, 0.004, 0.006, 0.008, 0.012, 0.016 and 0.02, where each subset was run initially with $10^7$ stellar binaries making a total of $1.2\times10^{8}$ initial binary systems.
For each metallicity we get a merging compact object catalogue for BNSs, BHNSs, and BBHs, containing properties such as delay times and masses of compact objects.

The \eagle\ suite \citep[][]{Schaye15,Crain15} is a set of hydrodynamical cosmological simulations run with a modified version of the code {\sc gadget-3}.
It includes subgrid models that account for various physical processes behind galaxy formation and evolution such as star formation, chemical enrichment, radiative cooling and heating, SN feedback, AGN feedback, and UV/X-ray ionizing background. 

The \eagle\ suite was run from $z = 127$ to $z \sim 0$, adopting $\Lambda$CDM scenario with cosmological parameters from \cite{Planck13} ($\Omega_{\rm m} = 0.2588$, $\Omega_\Lambda = 0.693$, $\Omega_{\rm b} = 0.0482$, and $H_0 = 100\,{} h$ km s$^{-1}$ Mpc$^{-1}$ with $h = 0.6777$).
The catalogues from the \eagle\ suite are available in the {\sc sql} database\footnote{{\url{http://icc.dur.ac.uk/Eagle/}}, {\url{http://virgo.dur.ac.uk/}}.}.

We use the galaxy catalogue from {\sc RefL0100N1504}. This run represents a periodic box of $100\Mpc$ side which initially contains $1504^3$ gas and dark matter particles of mass $m_{\rm gas} = 1.81\times10^6 \Msun$ and $m_{\rm DM} = 9.70\times10^6 \Msun$ (henceforth we refer to this run simply as \eagle{}). From the database, we extract galaxy properties such as stellar mass $M_\ast{}$, SFR, and metallicity of star-forming gas. We also get the properties of stellar particles formed in each simulated galaxy (i.e., mass, metallicity and formation time).

We implement a Monte Carlo algorithm to populate galaxies with merging compact objects generated with {\sc mobse} by assigning a number of merging systems to each stellar particle in the \eagle{} catalogue, according to their initial stellar mass and metallicity.
We assume that the formation time of the compact binary progenitors is equal to the formation time of the \eagle{} stellar particle to which they were assigned. We also assume that each binary compact object merges in the \eagle{} stellar particle where it was originally placed.
With this methodology, we obtain a population of merging compact objects for each galaxy in the \eagle{} catalogue across cosmic time.

We investigate the host galaxies of BNSs, BHNSs, and BBHs at four redshift intervals corresponding to $z=0-0.1$, $z=0.93-1.13$, $z=1.87-2.12$, and $z=5.73-6.51$. 
For simplicity, we refer to these intervals as $z=0.1$, $z=1$, $z=2$ and $z=6$ respectively. Thus, we map different fundamental stages:  the local Universe ($z\leq{}0.1$), the advanced LIGO and Virgo horizon for BBHs at design sensitivity ($z\sim{}1$), the peak of the cosmic star formation ($z\sim{}2$), and near the end of the cosmic reionization epoch ($z\sim{}6$).
Due to the resolution of the \eagle{} simulation, our analysis is focused on galaxies with  $M_\ast{} \ge{} 10^7~\Msun$. The selected galaxies also fulfill the condition ${\rm SFR} > 0\,{}\Msun{}$ yr$^{-1}$ (i.e., we remove from the sample the galaxies with ${\rm SFR} = 0\,{}\Msun{}$ yr$^{-1}$). In summary, the number of simulated galaxies selected from the \eagle\ simulation are 77959, 91294, 116074, and 50544 at $z=0.1$, 1, 2, and 6, respectively.

\subsection{Host-galaxy probability}\label{sec:method_prob}

In this work, we are interested in using the density of mergers per galaxy as a function of $M_{\ast{}}$, SFR and $Z$ to build a probability-like quantity called $p(M_{\ast{}}, {\rm SFR}, Z)$, that we will refer to as \textit{host-galaxy probability}. As shown later in this section, $p(M_{\ast{}},{\rm SFR},Z)$ can be used to derive the relative probability that a galaxy hosts a GW merger for a selected set of targeted galaxies, making this quantity particularly relevant for EM follow-up ranking procedure.

For each redshift bin, we approximate the values of the host galaxy probability as a piecewise constant function in bins of ($M_{\ast{}}$,~SFR,~$Z$) with values 
\begin{equation}\label{eq:hostp}
p^X_{i,j,k}=p_{\rm norm}\,{}\frac{N_{{\rm GW}\,{}i,\,{}j,{}k}^X}{N_{{\rm gal}\,{}i,\,{}j,{k}}},
\end{equation}
where $N_{{\rm GW}\,{}i,\,{}j,{k}}^X$  is the sum of all events of kind $X$ (where $X$ = BNS, BHNS or BBH) occurring in galaxies with masses in the $i$th bin (between $M_{\ast{},i}-\Delta{}M_\ast/2$ and $M_{\ast{},i}+\Delta{}M_\ast/2$), SFR in the $j$th bin (between ${\rm SFR}_j-\Delta{}{\rm SFR}/2$ and ${\rm SFR}_j+\Delta{}{\rm SFR}/2$) and $Z$ in the $k$th bin (between $Z_k-\Delta{}Z/2$ and $Z_k+\Delta{}Z/2$), while $N_{{\rm gal}\,{}i,\,{}j,{}k}$ is the sum of all galaxies with $M_{\ast{}}$ in the $i$th bin, SFR in the $j$th bin and $Z$ in the $k$th bin. Finally, $p_{\rm norm}$ is a normalization factor chosen so that $\sum_{i,j,k} p^X_{i,j,k} = 1$.

By marginalising over the metallicity, one can obtain the host galaxy probability $p(M_{\ast{}},{\rm SFR})$ taking values
\begin{equation}
p^X_{i,j}= \sum_{k=1}^{N_{\rm Z} } p^X_{i,j,k},
\label{binned_eq_proba}
\end{equation}
where $N_{Z}$ is the number of bins along the metallicity axis. Similarly, we can marginalise a second time over SFR and $M_{\ast{}}$ to obtain respectively the host galaxy probabilities $p(M_{\ast{}})$ and $p$(SFR).

Following a similar approach to \citet{Artale2019b}, we use a grid of $25\times25\times25$ points, with a range of $\log(M_\ast{} / \Msun) \in{}[ 7.0,\,{}12.5]$, $\log({\rm SFR} / \Msun \yr^{-1}) \in{}[ -5.5,\,{} 2.5]$, and $\log Z \in{} [-8,-0.7]$, in order to compute the elements of the host galaxy probability $p(M_{\ast{}},{\rm SFR},Z)$ in Eq.~\eqref{binned_eq_proba}.

In practice, the host galaxy probability that we have constructed can be directly integrated inside ranking procedures used by EM follow-up searches. To illustrate this, let us take an example where a GW detection with $N_{G}$ candidate galaxies identified and for which we have access to the stellar mass and SFR. If we take the very simplistic case where we both assume that all possible galaxies have been identified and have the same sky localisation probabilities, the relative probability that the $k-$th galaxy of our sample is the host of the merger is given by 
\begin{equation}
\mathbb{P}(\text{galaxy }k) = \frac{p(M_\ast{}^{k},\text{SFR}^{k} )}{\sum_{i=1}^{N_{G}} p(M_\ast{}^{i},\text{SFR}^{i} )} 
\label{proba_ranking}
\end{equation}
where ($M_\ast{}^{k}$,$\text{SFR}^{k}$) are the stellar mass and SFR of the $k-$th galaxy of our targeted search sample and $p(M_\ast{}^{k},\text{SFR}^{k} )$ is computed using Eq. \eqref{binned_eq_proba} with the appropriate bin.  If we now add the information coming from the sky localisation of the source,
  the relative probability that the $k-$th galaxy of our sample is the host of the merger is given by 
\begin{equation}
\tilde{\mathbb{P}}(\text{galaxy }k) =  \dfrac{\mathbb{P}(\text{galaxy }k)\,{}\,{}\mathbb{P}_{\text{loc}}(\text{galaxy }k)}{\sum_{k} \mathbb{P}(\text{galaxy }k)\,{}\,{}\mathbb{P}_{\text{loc}}(\text{galaxy }k)} ,
\label{proba_ranking2}
\end{equation}
where $\mathbb{P}(\text{galaxy }k)$ is the quantity defined in eq.~\ref{proba_ranking} and $\mathbb{P}_{\text{loc}}(\text{galaxy})$ is the probability that the galaxy hosts the merger based on sky localisation and redshift probability. Furthermore, the previous equation can easily be updated in more complex analysis that takes into account other factors such as the integration of galaxies without mass information or the effect of the incompleteness of the galaxy catalogue used (e.g. \citealt{Ducoin2019}).

  Using equation~\ref{proba_ranking2}, we can assign  a probability $\tilde{\mathbb{P}}(\text{galaxy }k)$  to each galaxy of our catalogue identified in the error box of a GW event. If the EM counterpart is not found, each galaxy in the box is at least ranked by this probability. This prioritization  can be used to give a different weight to each galaxy when, e.g., we want to estimate the Hubble constant $H_0$ from GW data. Namely, our  probability can be folded in equation~5 of \cite{abbottH02019} to refine our knowledge of the host galaxy distance.

Finally, we highlight that the host-galaxy probability defined in Eq.~\eqref{eq:hostp} is significantly different from the merger probability that was calculated in Eq.~(4) of \cite{Artale2019b}. The merger probability discussed in \cite{Artale2019b} accounts for two different ingredients: i) how common a galaxy with a given stellar mass and SFR at a given redshift in the Universe is (massive galaxies are way less numerous than low-mass galaxies); ii) the dependence of the number of compact-binary mergers on the stellar mass and SFR of the host galaxy. In contrast, the probability defined here in Eq.~\eqref{eq:hostp} accounts only for the second ingredient.

\section{Results}\label{sec:results}

In Section~\ref{sec:results_intrinsicMs}, we discuss the marginalized host galaxy probability as a function of the stellar mass, $p(M_{\ast{}})$,  at redshift $z=0.1, 1, 2$ and 6. The dependence on both SFR and  $M_{\ast{}}$ is treated in Section~\ref{sec:results_intrinsic_Ms_SFR}.

\subsection{Host-galaxy probability: stellar mass}\label{sec:results_intrinsicMs}

\begin{figure*}
\includegraphics[width=1.8\columnwidth]{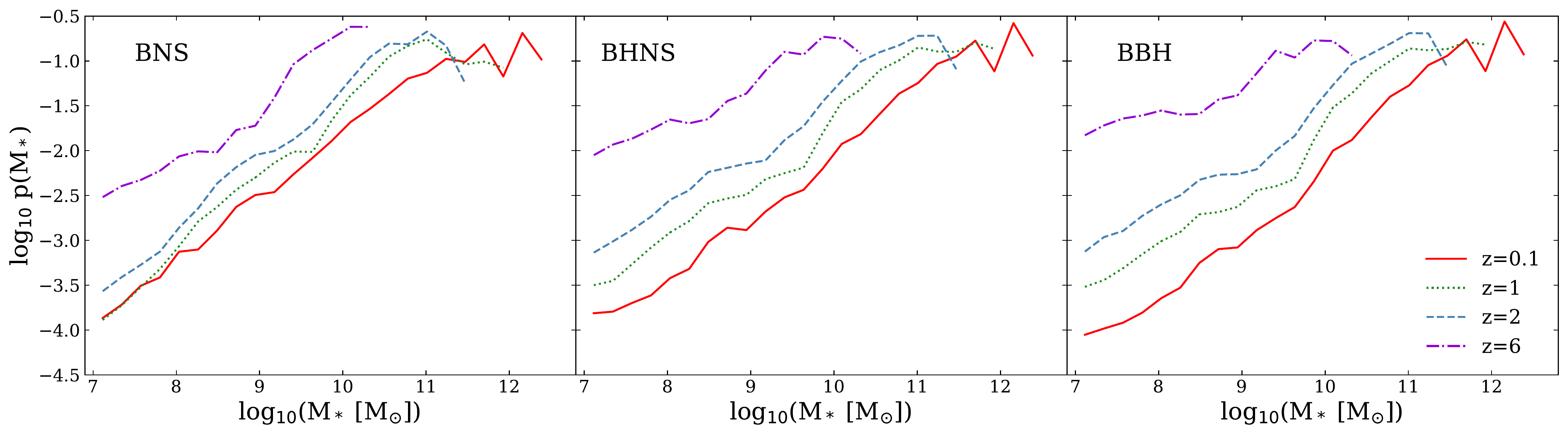}
\caption{Host galaxy probability $p(M_{\ast{}})$ as a function of stellar mass for the host galaxies of merging BNSs, BHNSs, and BBHs (from left to right) at $z=0.1$ (red solid lines), 1 (green dotted lines), 2 (blue dashed lines), and 6 (purple dashed-dotted lines).}
\label{fig:1D-HostGalaxyProb}
\end{figure*}

Figure~\ref{fig:1D-HostGalaxyProb} shows the host galaxy probability as a function of stellar mass, $p(M_{\ast{}})$, for merging BNSs, BHNSs, and BBHs, at redshifts $z=0.1, 1, 2,$ and 6. In all cases, massive galaxies are more likely to host merging compact objects than low mass galaxies. In addition, the peak of the distributions shifts towards higher values of stellar mass for decreasing values of redshift, as expected from galaxy assembly history (i.e., galaxies accrete more mass and become more massive with time). 
 Moreover, the host-galaxy probability for BBHs and BHNSs has a stronger dependence on redshift than for BNSs. 

These features are explained by the differences in the delay time distributions of the merging compact objects \citep[see][]{Mapelli2018,Mapelli2019}. In fact, BNSs have a delay time distribution that approximately scales with $\sim t^{-1}$ over the entire redshift range studied here, while merging BHNSs and BBHs  tend to have a flatter delay time distribution at $z \le{} 0.1$. 

The host probability associated with the smallest galaxies ($M_\ast\lesssim{}10^8$ M$_\odot$) is significantly higher at high redshift ($z=6$). This is another effect of delay time, as only compact binaries with a very short delay time merge by $z=6$, because the Universe at $z\sim{}6$ was younger than 1~Gyr and the first stars formed only few hundreds of Myr before. Hence, most of the compact binaries which merge at $z\sim{}6$ form and merge in the same galaxy. 

The host probability as a function of mass increases almost monotonically from $M_\ast{}=10^7$ M$_\odot$ to $\sim{}10^{10}$ M$_\odot$, but then it shows a turn-over at $M_\ast\ge{}10^{10}$ M$_\odot$. The turn-over point shifts toward higher masses as redshift decreases. Low-statistics effects affect our results at the high-mass tail: due to the size of the box of the {\sc{EAGLE}}, $\le{}10$ galaxies have stellar mass $\log (M_{\ast{}}/M_\odot) \ge{} 11.69,$ 11.46, and 11.12 at $z=0.1,$ 1, and 2, respectively. The most evident effect of low-statistics  is the oscillatory behaviour of the distribution for $\log (M_{\ast{}}/M_\odot) \ge{} 11.7$ at $z=0.1$. 

\begin{figure*}
\includegraphics[width=1.8\columnwidth]{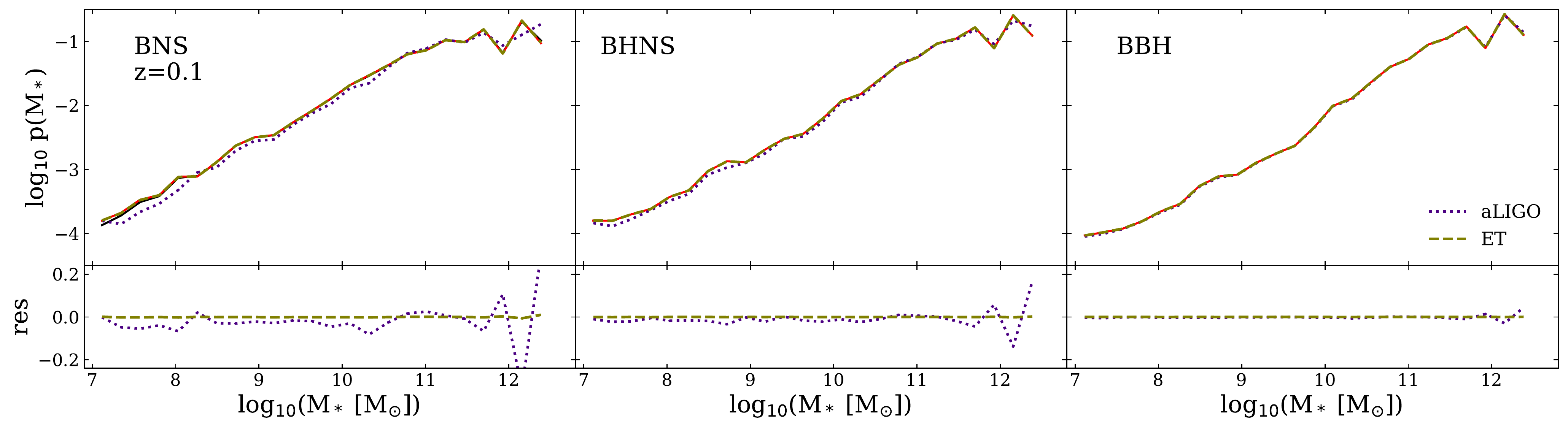}
\includegraphics[width=1.8\columnwidth]{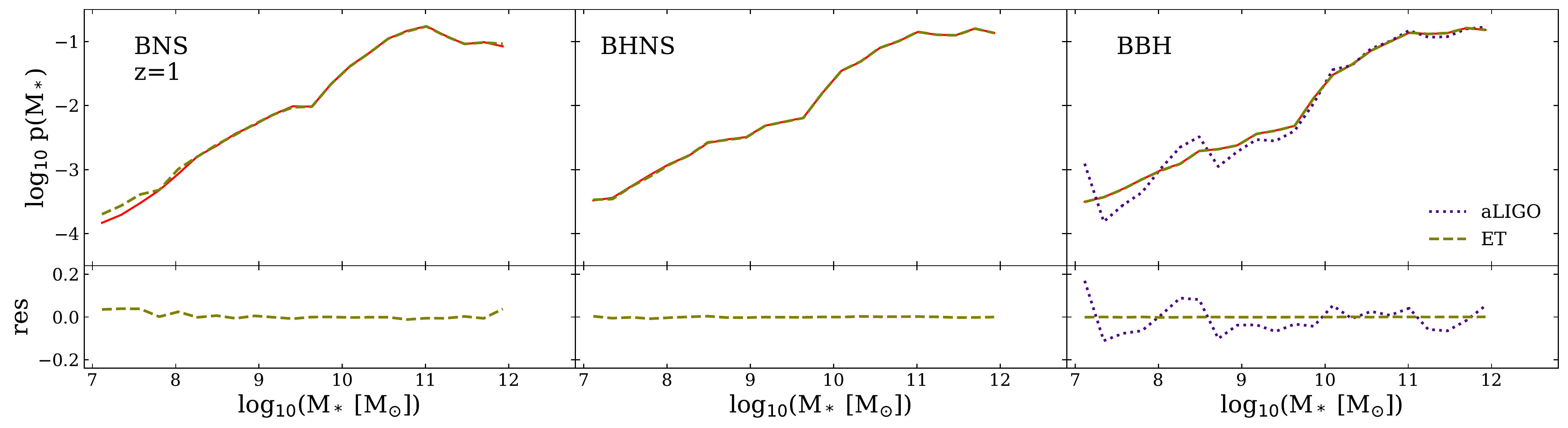}
\caption{Host galaxy probability as a function of stellar mass, for the downsampled set without detectability effects (red solid line), and with detectability effects, considering advanced LIGO-Virgo design sensitivity (purple dotted line), and ET (green dashed line) at redshifts $z=0.1$ (top) and $z=1$ (bottom). The bottom panel of each cell shows the residuals between the host galaxy probability without detection effects and the one with detection effects for advanced LIGO-Virgo (purple dotted line) and ET (green dashed line).}
\label{fig:1D-HostGalaxyProb-Det}
\end{figure*}

So far, our analysis directly took the distributions predicted by our model without taking into account detectability effects. However, we know that the current and future GW detectors introduce a bias in the observed distributions of compact mergers that should directly impact the evaluation of the host galaxy probability. To account for this, we have created catalogues of observed compact binary mergers by filtering our original catalogue using the framework already presented in previous works \citep{finn1992,dominik2015,Chen2018,Gerosa2018}, and summarised in Appendix~\ref{sec:appendix}. We highlight here that, given the computational cost of this analysis and the high number of binaries contained in the original catalogue (for instance, we have more than $6\times10^{8}$ BNSs at $z=1$), all the catalogues were downsampled in order to have a number of sources close to $10^{7}$. This number represents $\sim 1-20\%$ of the original intrinsic catalogues, and allows us to have manageable computation time. To check the validity of our downsampling procedure, we compared the values for $p(M_{\ast{}})$ obtained with the full set and the downsampled one. We found a maximal percentage error value close to $3 \%$, indicating very little bias between the two sets.

Figure~\ref{fig:1D-HostGalaxyProb-Det} shows the values of $p(M_{\ast{}})$ inferred from the downsampled catalogues 
that include detectability effects, considering advanced LIGO-Virgo design sensitivity 
and Einstein Telescope (hereafter ET)
at $z=0.1$, and 1. 
BNS and BHNS sources are not detected by advanced LIGO at $z=1$ due to the lack of sensitivity of the instrument at this redshift value.

At $z=0.1$, including detectability effects does not change the general features of $p(M_{\ast{}})$, that still favours galaxies with high mass. However, we do observe some bias introduced by detectability effects for advanced LIGO-Virgo, especially at high values of mass $M_{\ast{}}> 10^{11} M_{\odot}$. We find values of residuals as high as 0.25 for BNSs, while the maximum value of the residuals peaks at $~0.15$ and $0.02$ for BHNSs and BBHs, respectively (see Figure~\ref{fig:1D-HostGalaxyProb-Det}). This trend is explained by the fact that BNSs at redshift $z\sim{}0.1$ are already close to the instrumental horizon of advanced LIGO-Virgo, resulting in a loss of $~77\%$ of the sources after applying our filtering technique. For ET, we observe no differences between the two distributions and the residuals are almost always equal to zero. Once again, this is easily understood given that an ET-like detector can detect almost all compact binary mergers in our local Universe.

At $z=1$, $p(M_\ast{})$ does not change significantly when we consider the sensitivity of an ET-like observatory.  The situation is very different for advanced LIGO-Virgo, since only BBHs can be detected at this redshift, with a loss of $\sim 88\%$ of the intrinsic number of sources, resulting in a maximum value for the residual close to 0.2. 

\subsection{Host-galaxy probability: stellar mass and SFR}\label{sec:results_intrinsic_Ms_SFR}

\begin{figure*}
\includegraphics[width=\columnwidth]{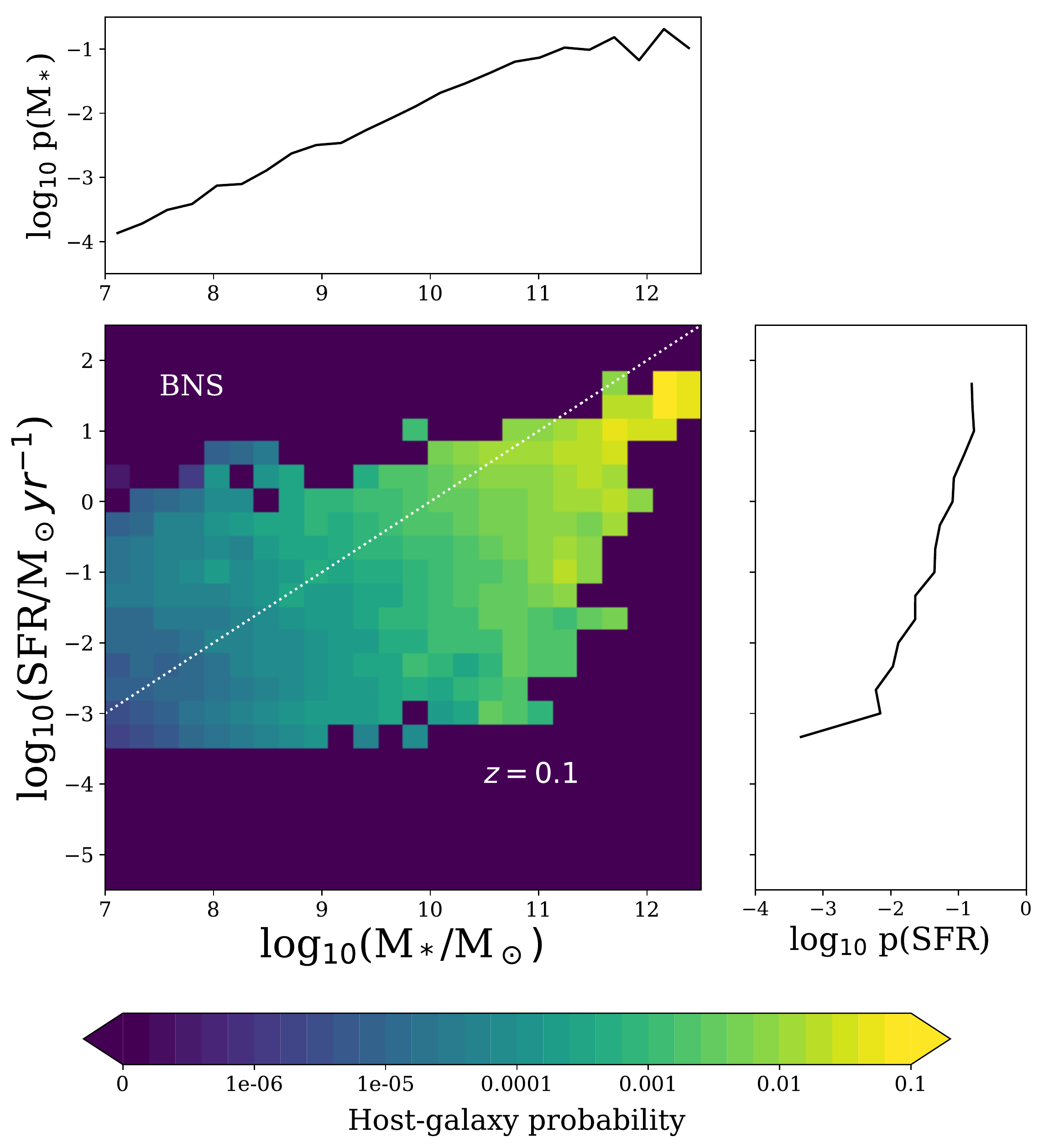}
\includegraphics[width=\columnwidth]{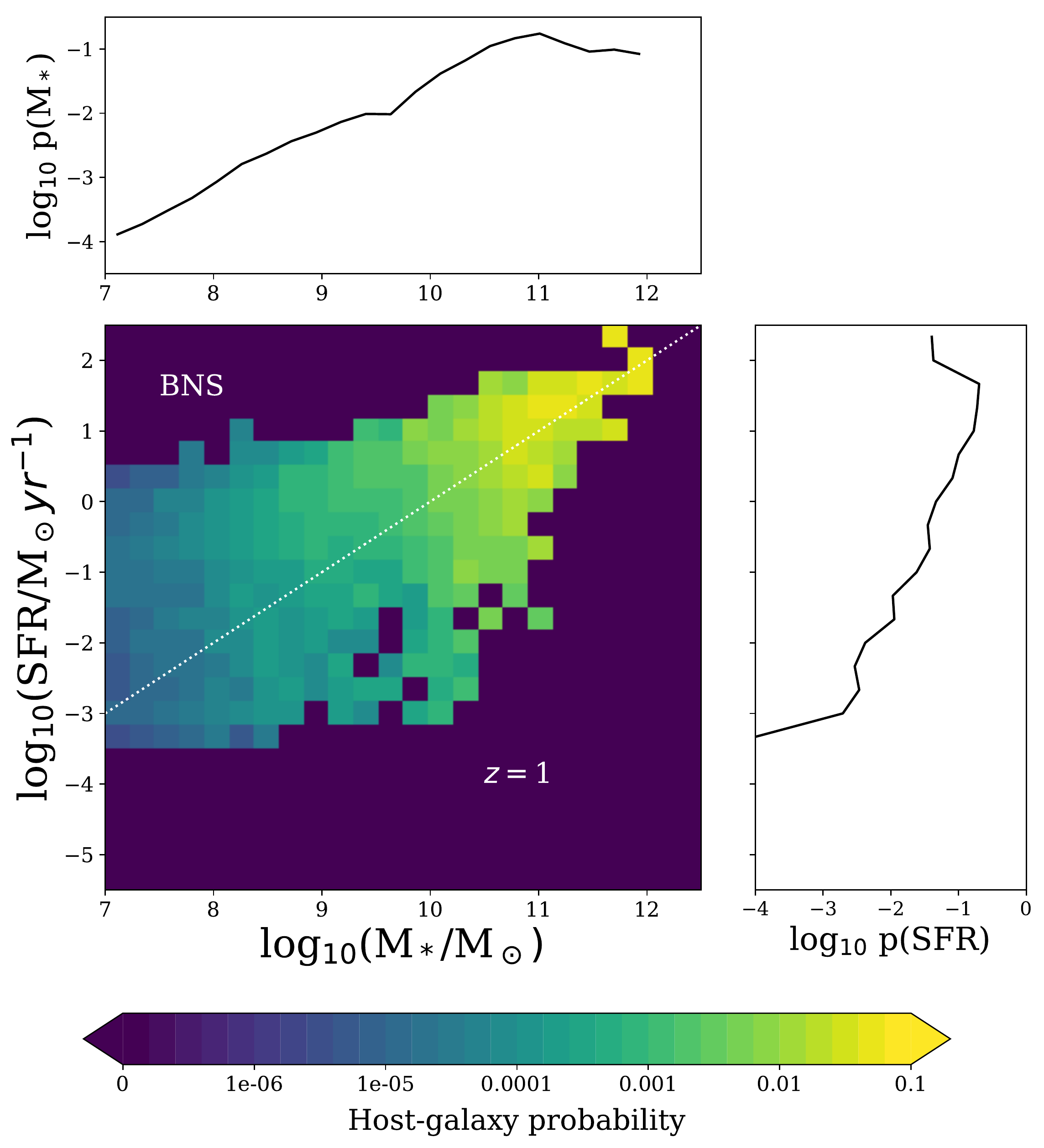}
\includegraphics[width=\columnwidth]{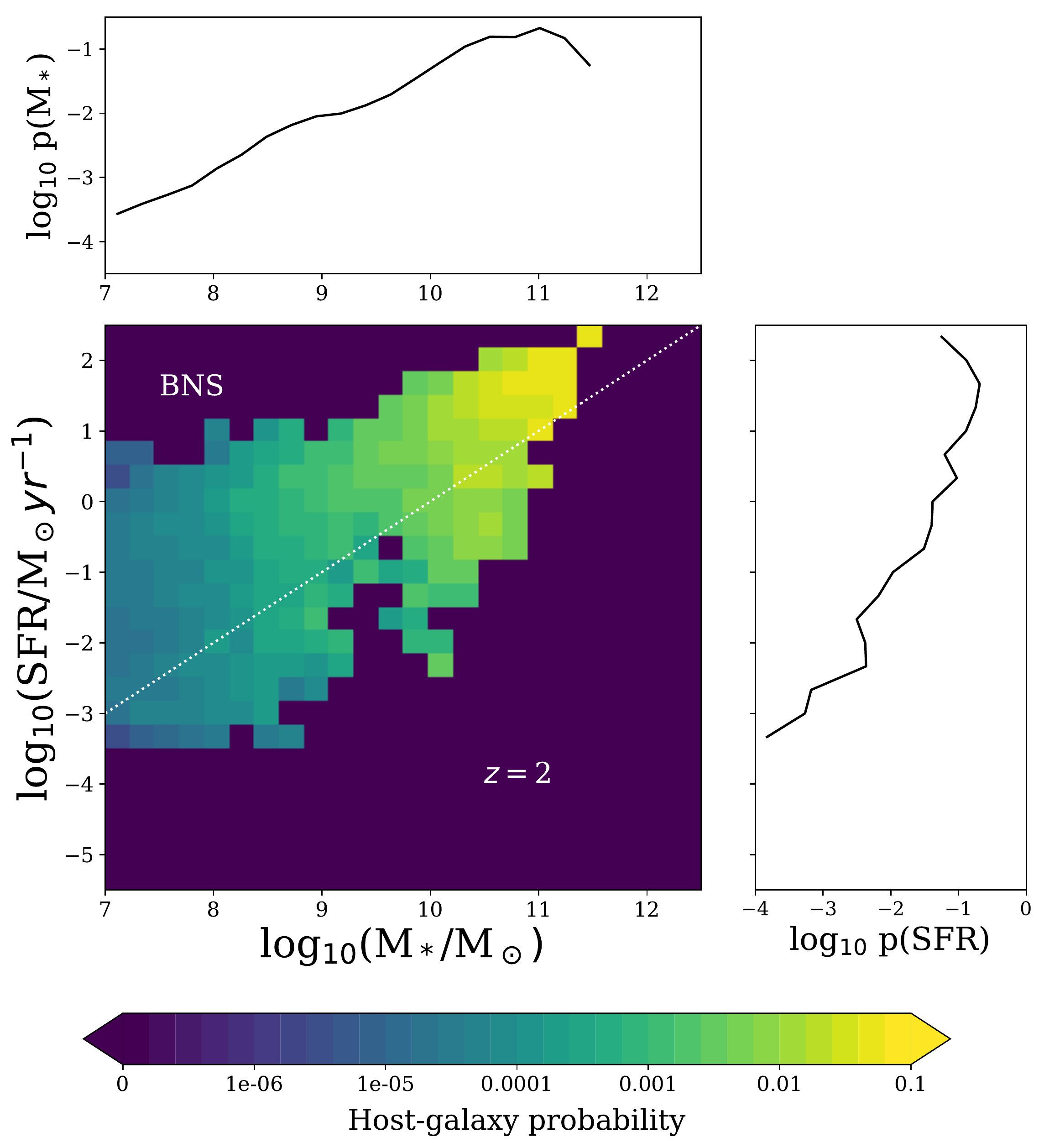}
\includegraphics[width=\columnwidth]{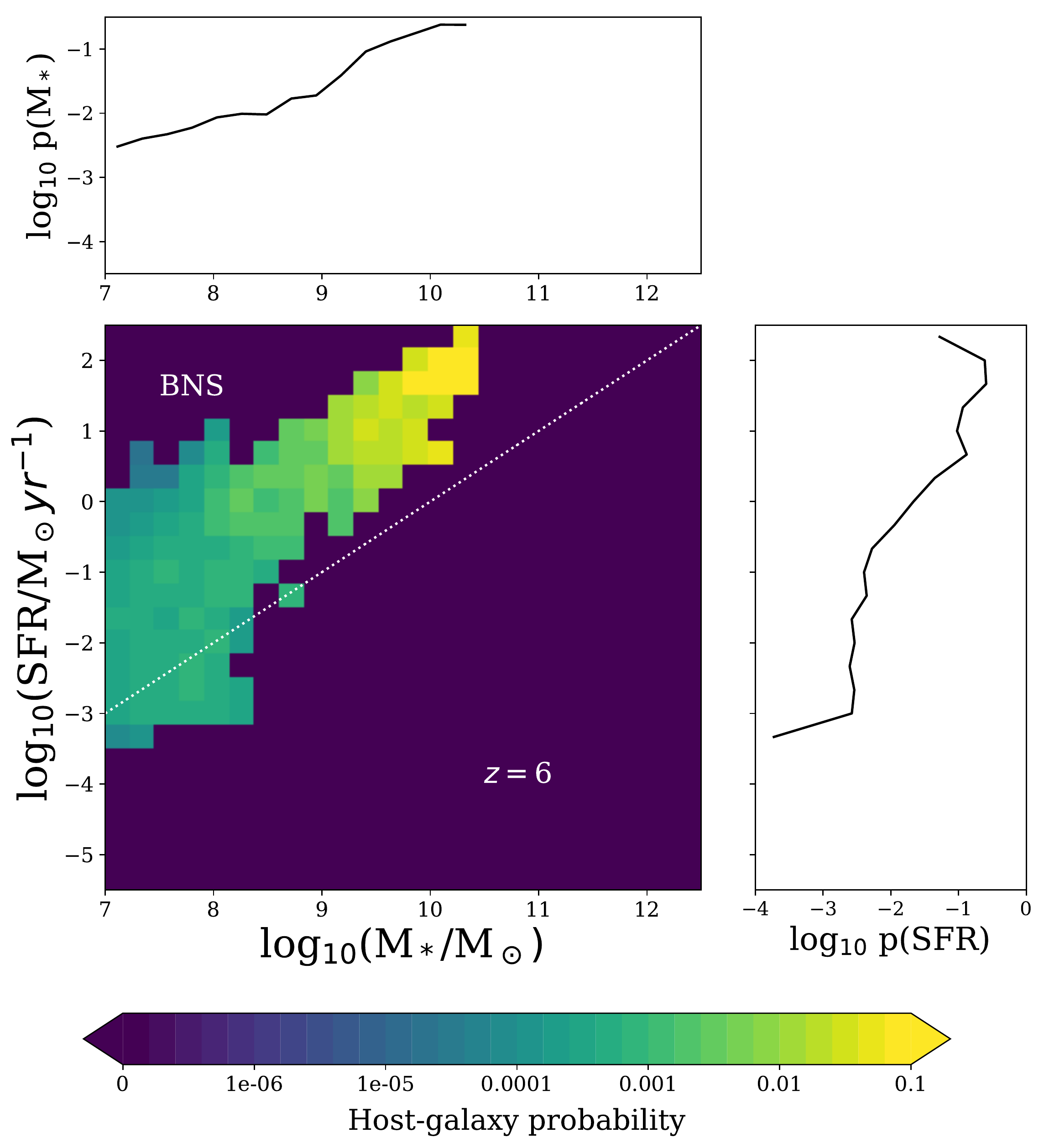}
\caption{Host-galaxy probability as a function of $M_\ast{}$ and SFR for the hosts of BNS mergers at redshifts $z = 0.1, 1, 2$ and 6.
The marginal histograms show the host galaxy probability as a function of SFR ($p({\rm SFR})$, right) and stellar mass ($p(M_\ast{})$, top). The white dotted line represents a specific SFR$ = 10^{-10} yr^{-1}$.}
\label{fig:DNS_prob_100Mpc}
\end{figure*}

\begin{figure*}
\includegraphics[width=\columnwidth]{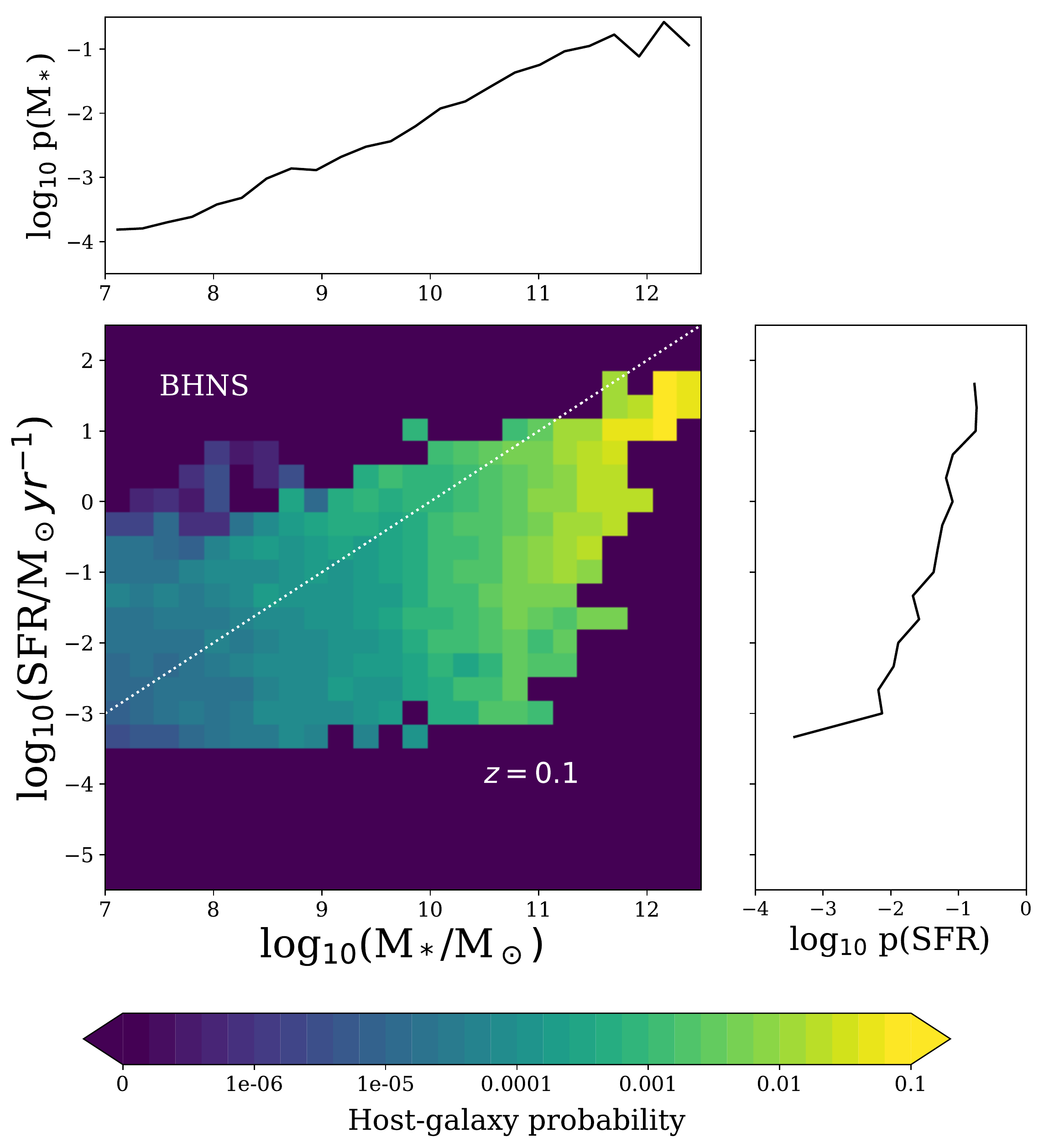}
\includegraphics[width=\columnwidth]{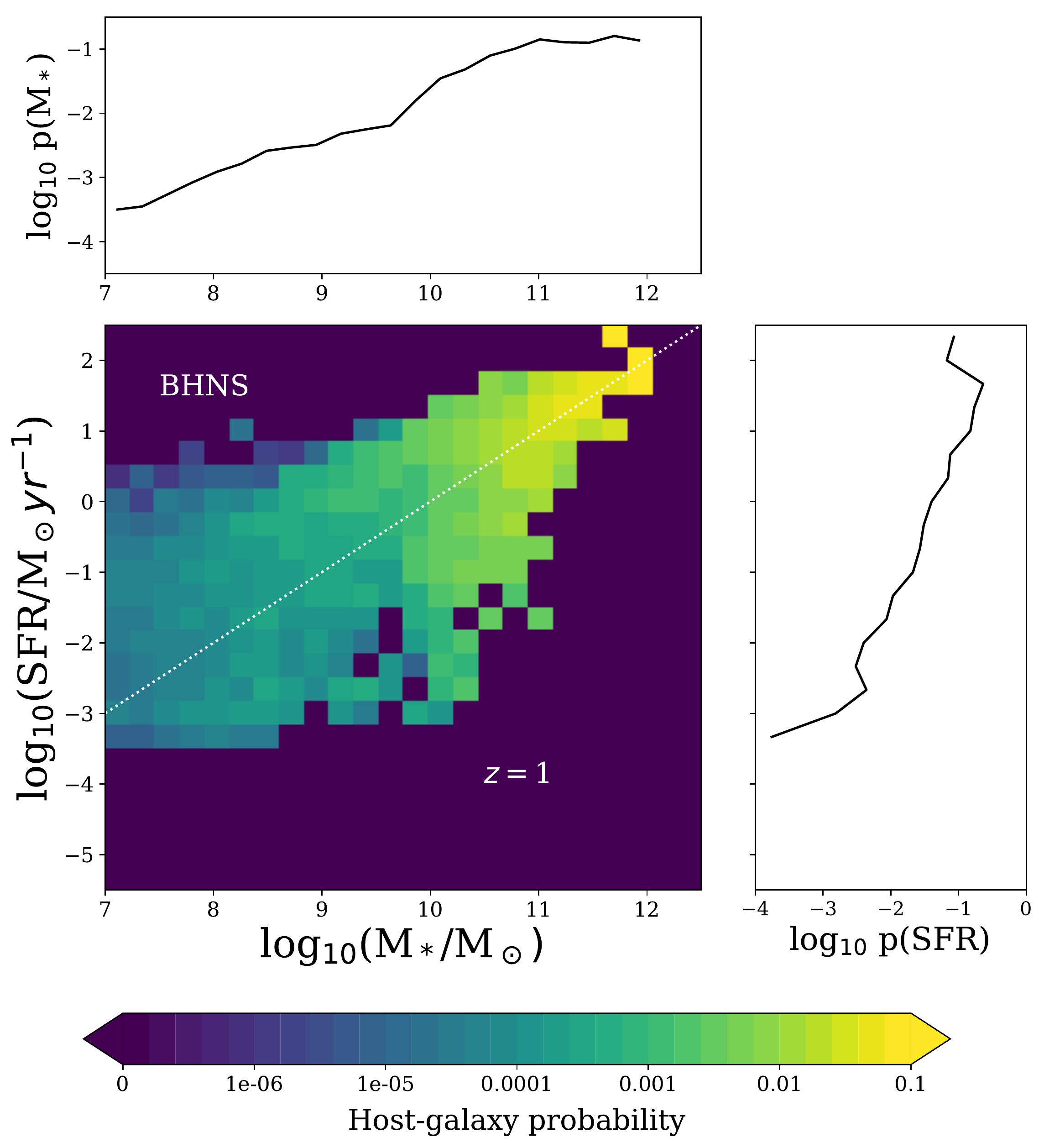}
\includegraphics[width=\columnwidth]{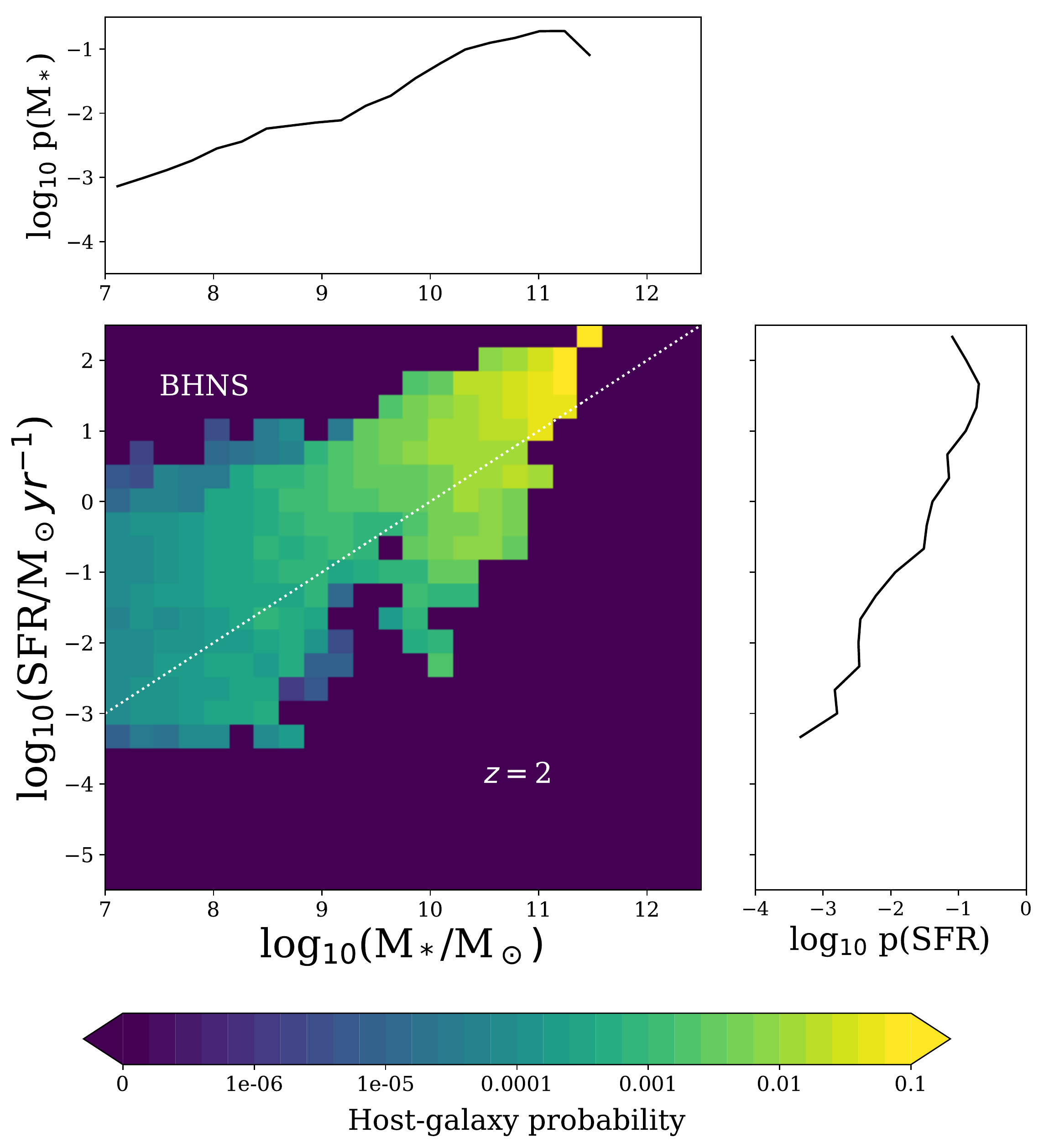}
\includegraphics[width=\columnwidth]{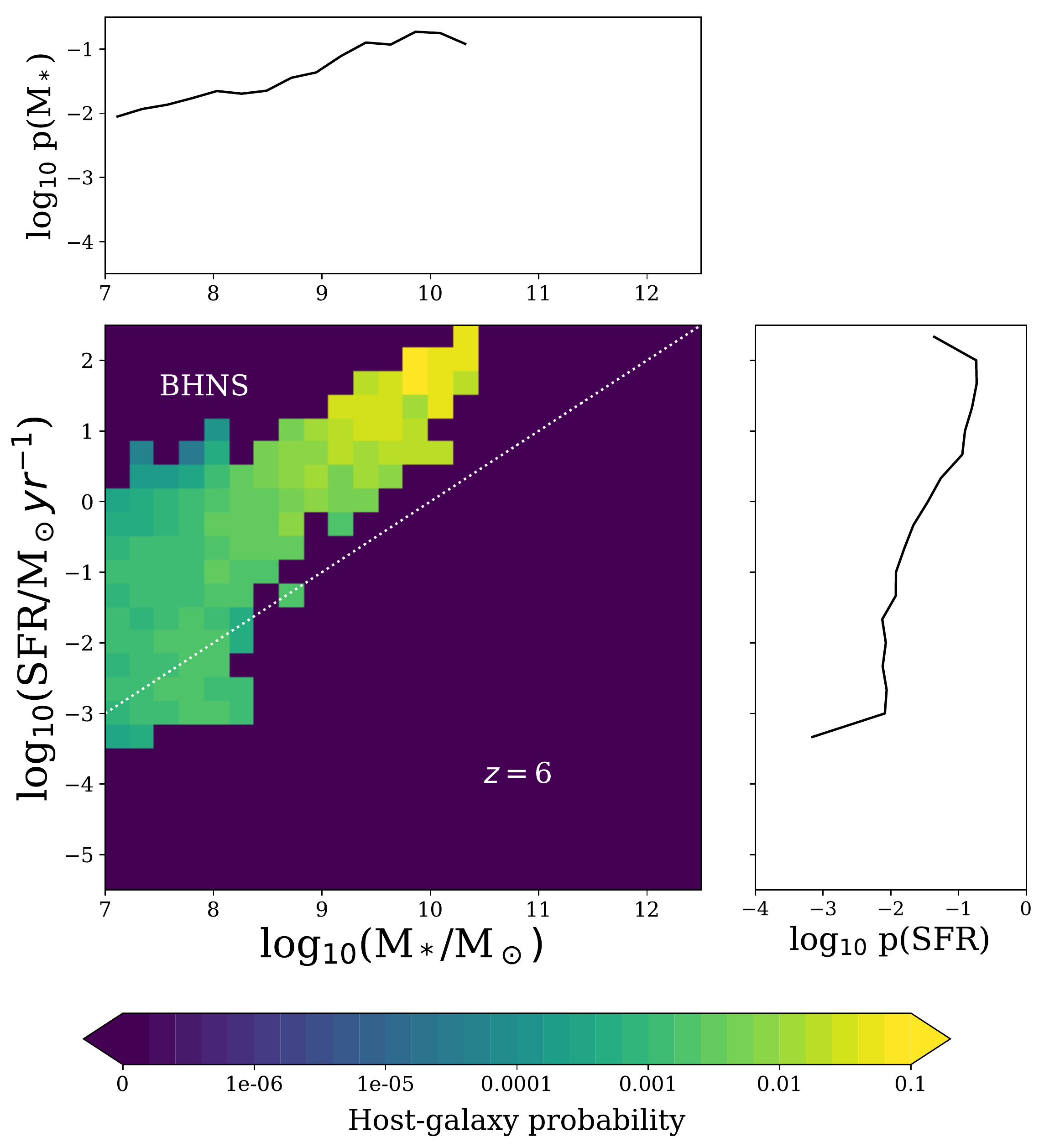}
\caption{The same as Figure~\ref{fig:DNS_prob_100Mpc}, but for the host galaxies of BHNSs.}
\label{fig:BHNS_prob_100Mpc}
\end{figure*}

\begin{figure*}
\includegraphics[width=\columnwidth]{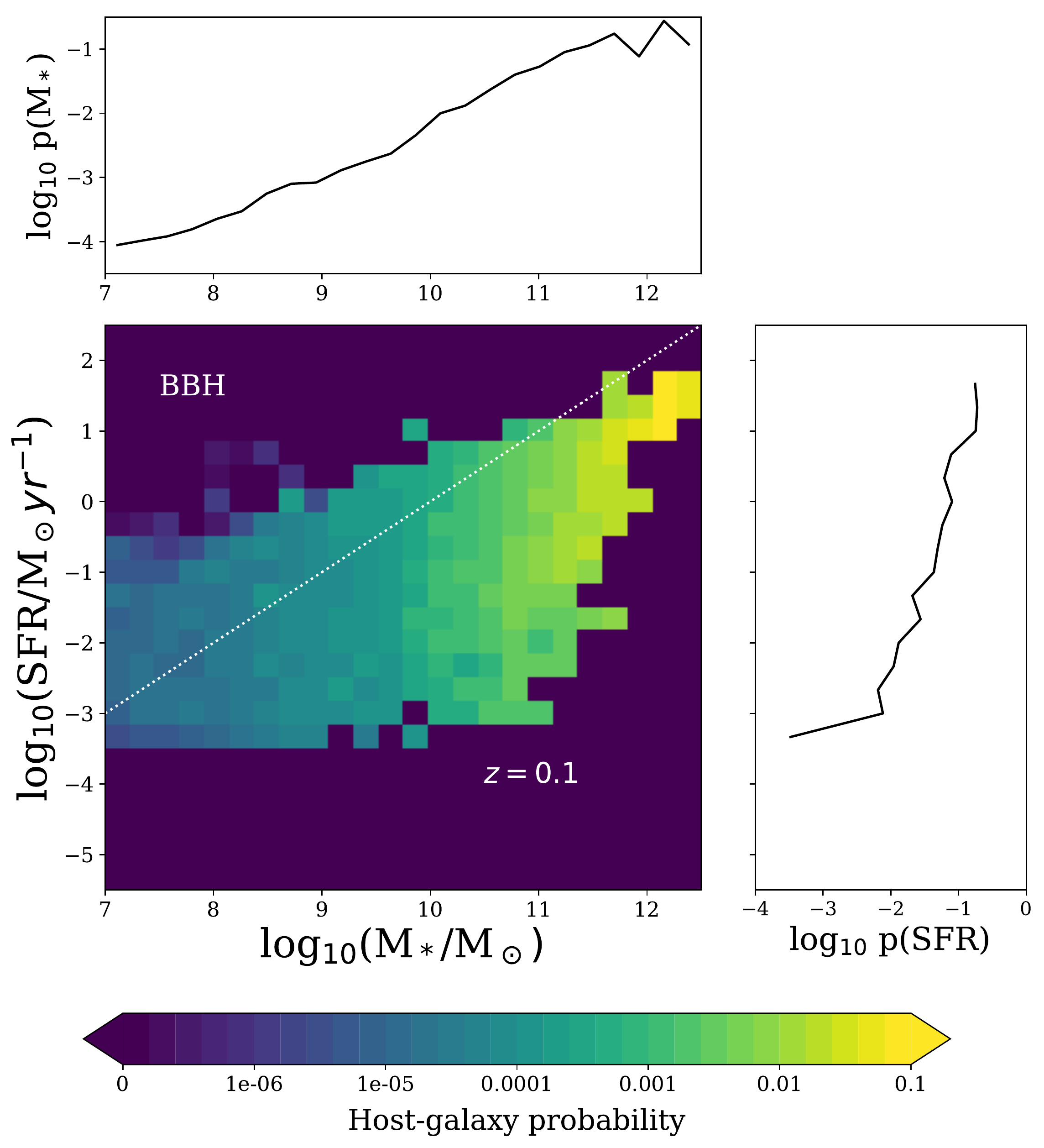}
\includegraphics[width=\columnwidth]{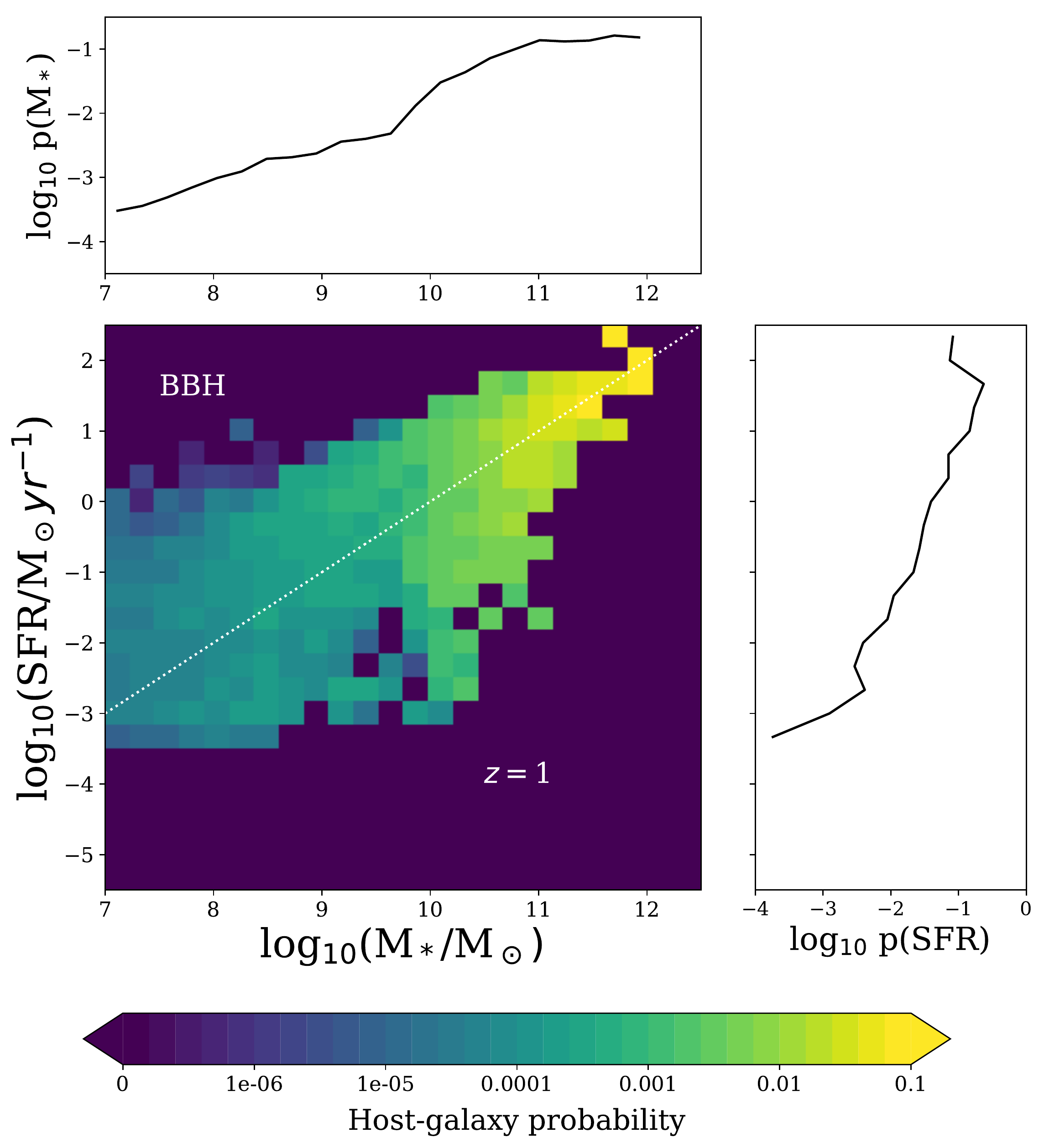}
\includegraphics[width=\columnwidth]{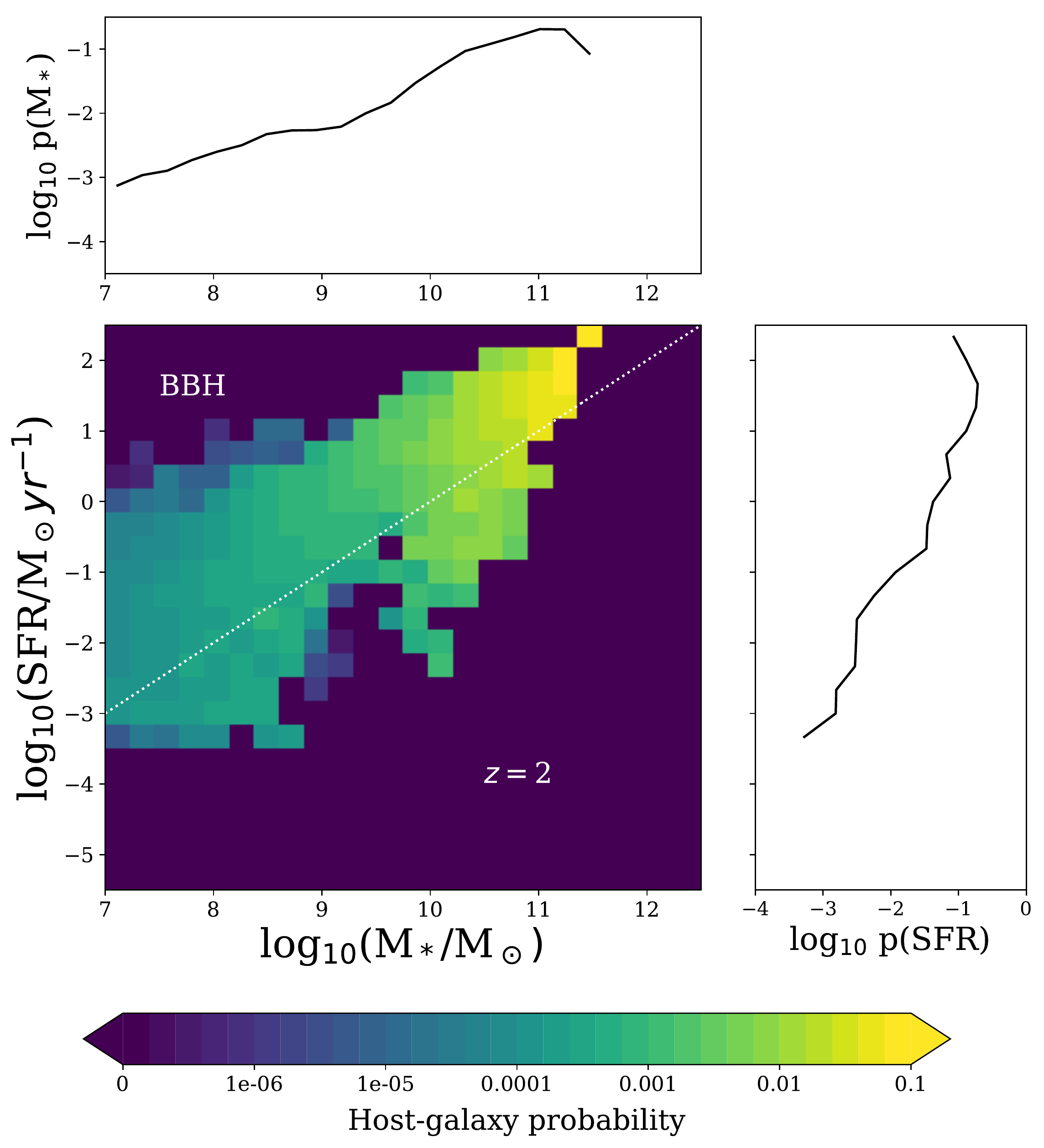}
\includegraphics[width=\columnwidth]{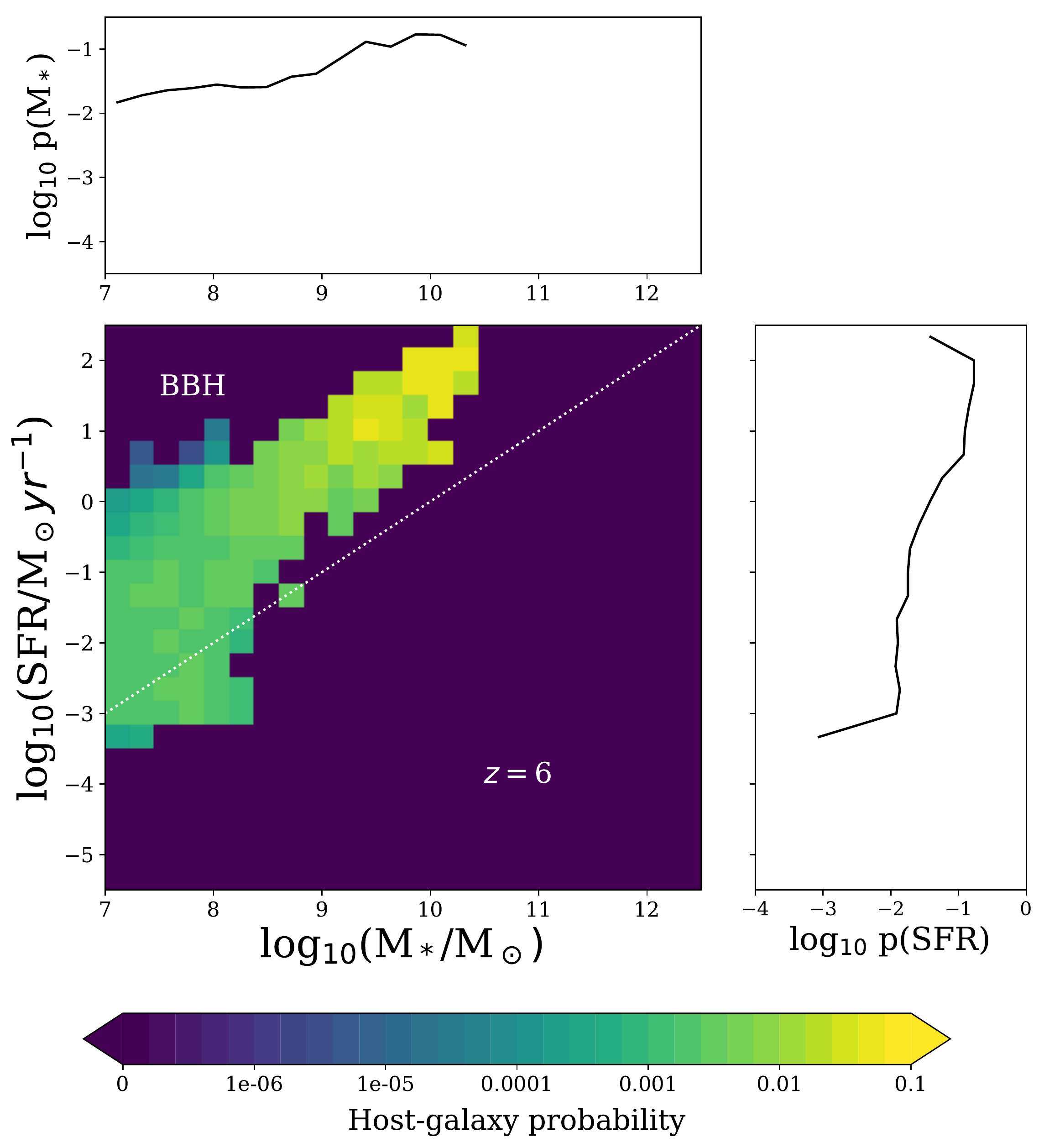}
\caption{The same as Figure~\ref{fig:DNS_prob_100Mpc}, but for the host galaxies of BBHs.}
\label{fig:DBH_prob_100Mpc}
\end{figure*}

SFR is another key ingredient that needs to be considered in the analysis. 
Several studies have shown that there is an intrinsic correlation between the stellar mass and the SFR of  galaxies \citep[see e.g.,][]{Furlong2015,Sparre2015,Trayford2019}. Figures~\ref{fig:DNS_prob_100Mpc},~\ref{fig:BHNS_prob_100Mpc} and ~\ref{fig:DBH_prob_100Mpc} show the values of the host-galaxy probability as a function of stellar mass and SFR, $p(M_{\ast{}},{\rm SFR})$, for merging BNSs, BHNSs, and BBHs, respectively, at $z = 0.1, 1, 2,$ and 6. These figures also display the marginal probability $p({\rm SFR})$ depending on SFR only.

Galaxies with higher values of SFR have a higher probability  $p(\text{SFR})$ to host a merger, regardless of the compact object type. 
If we now take a look at the probability $p(M_\ast,{\rm SFR})$ as a function of both stellar mass and SFR, we see that galaxies with high values of both stellar mass and SFR have a higher probability of hosting a merger, for all type of compact objects and for all considered redshifts. The shape of the  distributions also suggests that stellar mass is the main parameter influencing the host galaxy probability. In fact, for a fixed value of SFR, we observe more variation of the distribution than in the case where stellar mass is fixed. Quantitatively, we found that the ratio between the maximum and minimum (non-zero) value of the host galaxy probability is $10-100$ higher when SFR is fixed, compared to the case where $M_{\ast{}}$ is fixed. Hence, according to our model, the stellar mass of the host galaxies of compact binary mergers is more important than SFR to characterize the host galaxy probability.


\subsection{Host galaxy probability: $K_s$ and $B$ magnitudes}
Galaxy surveys used for the EM follow-up of GW sources (such as GLADE and CLU \citealt{Dalya2018,Cook2019}) do not include direct information on the stellar mass and SFR of galaxies. Rather, they 
contain $B-$band and $K_s-$band magnitudes as a proxy of SFR and stellar mass, respectively.

Hence, in order to facilitate the usage of our results in EM searches, we derive the host galaxy probability as a function of the rest-frame absolute magnitudes in $B-$ and $K_s-$band, $p(M_{\rm B})$ and $p(M_{\rm K_s})$ respectively. To this purpose, we have extracted the rest frame absolute magnitudes provided in the \eagle\ database. The magnitudes were computed for the galaxies with stellar mass above
$M_\ast{} \sim 1.81 \times 10^{8} {\rm M}_\odot$, and for the stellar emission including the effect of dust  \citep[see][for further details]{Trayford2019}. 

\begin{figure*}
\includegraphics[width=1.8\columnwidth]{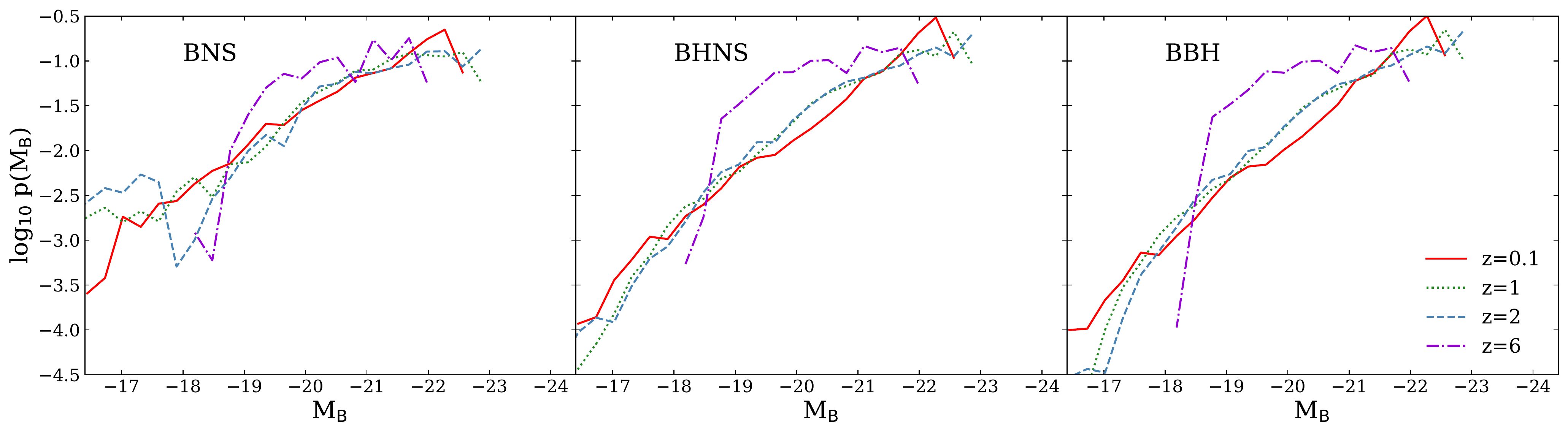}
\includegraphics[width=1.8\columnwidth]{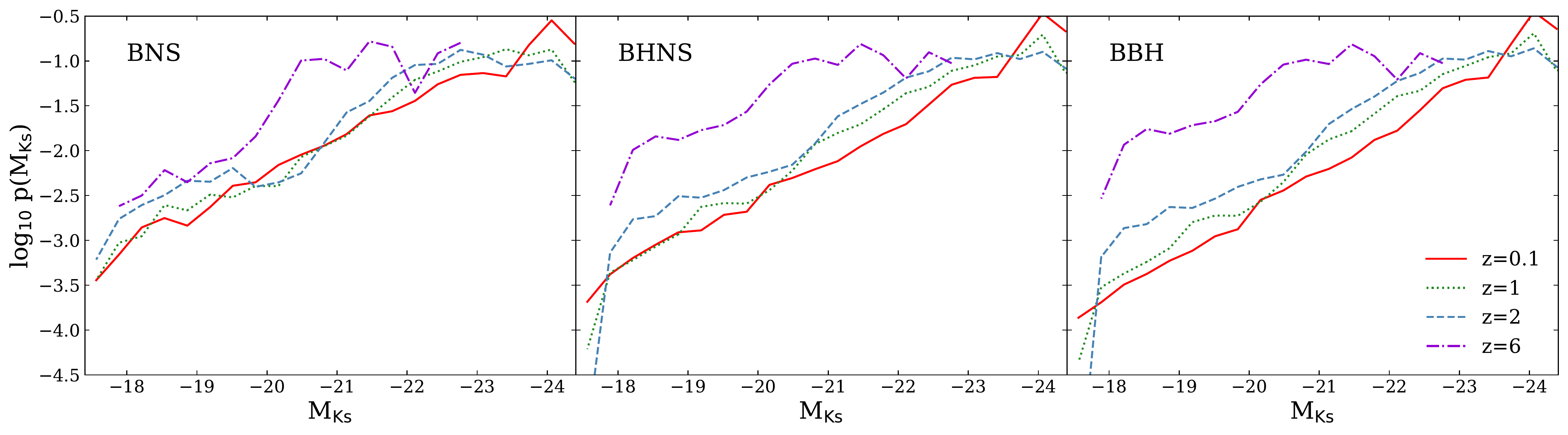}
\caption{Host galaxy probability as a function of the rest-frame absolute magnitudes in $B$ (top) and $K_s$ band (bottom) for the host galaxies of merging BNSs, BHNSs, and BBHs (from left to right) at $z=0.1$ (red solid lines), 1 (green dotted lines), 2 (blue dashed lines), and 6 (purple dashed-dotted lines).}
\label{fig:1DProb-Luminosities}
\end{figure*}

Figure~\ref{fig:1DProb-Luminosities} shows the host galaxy probability as a function of the rest-frame absolute magnitudes in $B-$ and $K_s-$band, $p(M_{B})$ and $p(M_{K_s})$. The host galaxy probabilities are computed with the same methodology as explained in Section~\ref{sec:method_prob}~and~\ref{sec:results_intrinsicMs}, but replacing the stellar mass of each galaxy with its rest-frame magnitude. Both $p(M_{B})$ and $p(M_{K_s})$ grow as a function of $M_{B}$ and $M_{K_s}$, respectively.

The peak of the host galaxy probability as a function of the $K_s$ magnitude shifts to higher luminosities (i.e., more massive galaxies) as redshift decreases, following the same trend that we found for $p(M_{\ast{}})$. This trend is stronger for the host galaxies of BHNSs and BBHs than for the host galaxies of BNSs.

In contrast, $p(M_{\rm B})$ does not show significant variations with redshift in the range $z=0.1-2$. Indeed, the lack of a trend with redshift is similar to what we found for the host galaxy probability as a function of the SFR $p({\rm SFR})$. These results confirm that while the $K_s$ magnitude is a good proxy for stellar mass,  the $B$  magnitude is affected by dust extinction, i.e., it is more a tracer of the SFR than a proxy for the stellar mass.

\section{Comparison to other prioritization strategies}\label{sec:discussion}
 Previous studies have developed prioritization strategies which include stellar mass \citep{Ducoin2019} or $B-$ magnitude \citep{Arcavi2017}. Our method is complementary to such strategies for the following reasons.
\begin{itemize}
\item{} Previous work assumes that the host-galaxy probability scales with the stellar mass (or with the $B-$band luminosity) based on heuristic arguments, such as the host-galaxy mass distribution of short gamma ray bursts \citep{Leibler2010, Fong2013, Berger2014}, and on the results of previous theoretical studies \citep{Artale2019b, Toffano2019, Mapelli2018}. Here, we directly compute a probability distribution from our astrophysical model. Hence, while our criterion is more model dependent than previous studies, it is also self-consistent and robustly motivated from an astrophysics point of view.

\item{} In previous work, the host galaxy probability is assumed to scale linearly with stellar mass (excluding possible corrections, such as the correction for galaxies without mass measurement introduced by \citealt{Ducoin2019}). In contrast, we do not need to assume a linear scaling. Rather, we assign specific weights to each stellar mass bin (and/or SFR bin), which are uniquely determined by our model. We show that there is a correlation between the merger rate per galaxy and the stellar mass, and this correlation is not simply a linear relation. Weighting this correlation is important to quantify deviations with respect to a linear scaling. Future detections of EM counterparts will help us to validate our criterion, giving us a feedback on our astrophysical model.

\item{} In our methodology, we can use not only the stellar mass but also the SFR: when not only the stellar mass is known but also the SFR, both data can be used to improve the prioritization.

\item{} We generalize our method to higher redshifts. This will become important for next generation ground-based GW detectors \citep{punturo2010,reitze2019,Kalogera2019,maggiore2020}.

\item{} \cite{Ducoin2019} derive stellar masses from data using a specific calibration, while we follow a complementary approach: since observational catalogues usually do not contain stellar masses but magnitudes, we use the cosmological simulations to translate stellar masses and SFR into $K_S$ and $B$ band, respectively. This can be done for other bands, making our criterion even more flexible. If the $K_S$ and/or the $B$ magnitude, the stellar mass and/or the SFR are available for the galaxies in the error box, our tabulated probabilities can be used and convolved with sky localization as described in equation~\ref{proba_ranking2}.

  For example, we  have convolved our ranking criterion with the stellar masses in the MANGROVE catalog, to check consistency with the results of  \cite{Ducoin2019}. We found that, using their eq.~4 to rank the possible host galaxies of GW170817, we obtain the same ranking for the first three galaxies of the catalogue, namely NGC~4993, IC~4197 and NGC~4970.
  \end{itemize}

\section{Conclusions}\label{sec:conclusions}
Understanding what are the properties of the most likely host galaxies of BNS, BHNS and BBH mergers is fundamental for the EM follow-up of GW detections. 
Here, we discussed the host galaxy probability, i.e. the probability that a galaxy hosts a merging compact object as a function of its stellar mass and SFR at redshifts $z=0.1, 1, 2,$ and 6. Our methodology combines the results from the population synthesis code \mobse\ \citep{Giacobbo2018,Giacobbo2018B} together with galaxy catalogues from the cosmological simulation \eagle\ \citep{Schaye15}. With a Monte Carlo approach, we populate each galaxy in the \eagle\ catalogue with BNSs, BHNSs and BBHs.

Our results show that there is a strong dependence of the host galaxy probability on the stellar mass, $p(M_\ast)$,
for all types of compact binaries: massive galaxies are more likely to host a compact-binary merger than low-mass galaxies. Moreover, the distribution shifts towards massive hosts as redshift decreases (see Fig.~\ref{fig:1D-HostGalaxyProb}), as a consequence of the cosmic galaxy assembly and the delay time distribution of compact binary sources.  

We also find that galaxies with higher SFR have a higher probability of hosting compact object mergers. Nevertheless, as shown in Fig.~\ref{fig:DNS_prob_100Mpc},~\ref{fig:BHNS_prob_100Mpc},~and~\ref{fig:DBH_prob_100Mpc},  the dependence on SFR is not as strong as the dependence on stellar mass. At a fixed stellar mass, $p(M_\ast{},\text{SFR})$ shows less significant variation compared to the case where SFR is fixed. Overall, the peak of the host probability at high SFR is mainly due to massive galaxies which in turn are forming more stars.

Our conclusions depend on the population-synthesis model and on the cosmological simulation. On the other hand, it is remarkable that both BNSs and BBHs show the same trend: their preferred host galaxies are the most massive ones and possibly the most star forming. This result indicates that the dependence on a specific population-synthesis model is not particularly strong, if we consider that the merger efficiency of BNSs marginally depends on metallicity in our models, while the merger efficiency of BBHs and BHNSs dramatically depends on progenitor's metallicity. The key ingredient here is the delay time distribution, which is not strongly affected by model assumptions \citep{Mapelli2019}. Our results rely on the choice of the {\sc{EAGLE}} simulation. In \cite{Artale2019b}, we have shown that our results are not significantly influenced by resolution, although we are not able to study galaxies with stellar mass $M_\ast<10^7$ M$_\odot$ and $>10^{12}$ M$_\odot$. The {\sc{EAGLE}} appears to match some of the most important observables, such as the mass--metallicity relation of galaxies \citep{DeRossi2017}, the colour--magnitude diagram at $z=0$ \citep{Trayford2015}, the evolution of the galaxy stellar mass function, and the cosmic SFR density evolution \citep{Furlong2015,Trayford2019}.

In the context of the EM follow-up of GW sources, our results can be implemented in ranking procedures applied for GW mergers detected by advanced LIGO-Virgo \citep{Singer2016}. The standard strategy consists in targeting galaxies within the skymap provided by the LVC using a complete galaxy catalogue such as GLADE \citep{Dalya2018}. Then, the galaxies are ranked adopting a criterion generally based on the 2D localization region, distance, and $B-$band magnitude \citep[see e.g.,][]{Arcavi2017,Coughlin2018,Rana2019}. Recently, \cite{Ducoin2019} proposed to use the stellar mass instead of the $B-$band magnitude in the ranking procedure.

Here, we propose a new astrophysically motivated criterion to compute the relative probability that a galaxy inside the LVC uncertainty box 
hosts a GW detection. If we calculate this host galaxy probability as a function of stellar mass only ($p(M_\ast)$), our results are in agreement with the analytic model from \citet{Ducoin2019}, and we find that high mass galaxies are much more likely to host mergers than low mass galaxies.

In addition, we also investigate what is the impact of including the SFR in our ranking criterion, which to our knowledge was never done before. While we find that the effect of SFR is less important than stellar mass, we observe that the probability that a galaxy hosts a merger is higher for high values of both stellar mass and SFR.

Galaxy catalogues usually do not include direct information on stellar mass and SFR. Hence, to facilitate the usage of our ranking criterion, we calculate the host galaxy probabilities $p(K_s)$ and $p(B)$ as a function of the $K_s-$band and $B-$band magnitudes. The former shows a similar trend to $p(M_\ast)$, while the latter is reminiscent of $p({\rm SFR})$, suggesting that $K_s$ and $B$ magnitudes are tracers of stellar mass and SFR, respectively.

Our host-galaxy probabilities  can be implemented in future low-latency searches, providing a useful criterion to rank possible host galaxies within the sky localization map. The tables of the host-galaxy probabilities can be obtained online in table format\footnote{See Appendix~\ref{sec:suplementary}, or alternatively  \url{https://github.com/mcartale/HostGalaxyProbability}}.

\section*{Acknowledgements}
We thank Walter Del Pozzo, Marica Branchesi, Stefan Grimm, Jan Harms, and Dario Rodriguez for their useful comments. MCA and MM acknowledge financial support from the Austrian National Science Foundation through FWF stand-alone grant P31154-N27
``Unraveling merging neutron stars and black hole -- neutron star binaries with population synthesis simulations''.
YB, NG, MM and FS acknowledge financial support by the European Research Council for the ERC Consolidator grant DEMOBLACK, under contract no. 770017. 
MS acknowledges funding from the European Union's Horizon 2020 research and innovation programme under the Marie-Sk\l{}odowska-Curie grant agreement No. 794393. MP acknowledges funding from the European Union's Horizon $2020$ research and innovation programme under the Marie Sk\l{}odowska-Curie grant agreement No. $664931$.  
We acknowledge the Virgo Consortium for making their simulation data available. The \eagle\ simulations were performed using the DiRAC-2 facility
at Durham, managed by the ICC, and the PRACE facility Curie based in France at TGCC, CEA, Bruy\`{e}res-le-Ch\^{a}tel.

\bibliographystyle{mnras}

\bibliography{Artale_GW}

\IfFileExists{\jobname.bbl}{}
{\typeout{}
\typeout{****************************************************}
\typeout{****************************************************}
\typeout{** Please run "bibtex \jobname" to optain}
\typeout{** the bibliography and then re-run LaTeX}
\typeout{** twice to fix the references!}
\typeout{****************************************************}
\typeout{****************************************************}
\typeout{}
}

\appendix

\section{Joint probability distribution according to the detectability of ground based interferometers}\label{sec:appendix}

To produce catalogues of sources as observed by either LIGO-Virgo or ET, we follow a similar approach as done in previous works 
\citep[see e.g.,][]{finn1992,dominik2015,Gerosa2018}. Here, we briefly summarise the main details of this approach.

To assess if a source is detected or not, we look at the value of its signal-to-noise ratio (SNR) given by, 
\begin{equation}
\rho = 4 \int_{0}^{\infty}\dfrac{\mid \tilde{h}(f) \mid^{2}}{S_{n}(f)},
\end{equation}
where $\tilde{h}(f)$ is the gravitational waveform in the Fourier domain and $S_{n}(f)$ is the one-sided noise power spectral density of the detector. In this study, we have used the design sensitivity for advanced LIGO-Virgo and ET, as given in \cite{Abbott2018} and \cite{evans2016}, respectively. For the waveform models, we adopted the IMRPhenomD model \cite{khan2015} for BBH coalescence and TaylorF2 for BNS and BHNS coalescences, where we set the spins equal to 0. The waveforms and detector sensitivities were both generated by the package PyCBC \citep{canton2014,usman2015}. 

As our catalogues only provide values for the masses and the redshift, the dependence of the waveform on other parameters such as the position in the sky and inclination, is represented by the parameter $\omega$. The SNR is then expressed as $\rho = \omega \times \rho_{\rm opt}$ where $\rho_{\rm opt}$ is the optimal SNR when the source is face-on and optimally oriented. The probability of detecting a source is then given by 
\begin{equation}
p_{\text{det}}(w) = 1 - F_{\omega}(\rho_{\rm thr}/\rho_{\rm opt}),
\label{prob_det_GW}
\end{equation} 
where $F_{\omega}$ is the cumulative distribution function of $\omega$. This function can easily be approximated by Monte Carlo methods such as the ones implemented in the package we used \citep{gwdet_ref}. For the SNR threshold, we take $\rho_{\rm thr}=8$, as it was shown to be a good proxy of more complex analysis involving a network of detectors.

In practice, we evaluate the optimal SNR $\rho_{\rm opt}$ for each source in our catalogue and then compute the detection probability as in eq.~\eqref{prob_det_GW}. We then draw a random number $\epsilon \sim \mathcal{U}[0,1]$ and add the source with probability $p_{\text{det}}$ in our catalogues of detectable sources that are used to produce Fig.~\ref{fig:1D-HostGalaxyProb-Det}.

\section{Supplementary material: Host-galaxy probability tables}\label{sec:suplementary}

The host-galaxy probabilities for BNSs, BHNSs, and BBHs at $z=0.1, 1, 2$, and 6 presented in Fig.~\ref{fig:DNS_prob_100Mpc}, \ref{fig:BHNS_prob_100Mpc}, and \ref{fig:DBH_prob_100Mpc} can be obtained in table format in the supplementary material online. 
Alternatively, the tables are also available on \url{https://github.com/mcartale/HostGalaxyProbability}.
Our results can be directly implemented in low-latency searches of EM counterparts as described in Section~\ref{sec:method}.

\begin{table*}
\caption{Table of the host galaxy probability as a function of $M_\ast{}$ and SFR, for BNSs at $z=0.1$, corresponding to the results presented in Fig.~\ref{fig:DBH_prob_100Mpc}. Stellar masses are in units of M$_{\odot}$, while SFR are in yr$^{-1}$ M$_{\odot}$.
Note: This table is presented in its entire version in \url{https://github.com/mcartale/HostGalaxyProbability}. An abbreviated version is shown here for guidance.}
\begin{tabular}{|c|c|c|c|c|}
\cline{1-5}
$\log(M_\ast{}^{\rm min})$ & $\log(M_\ast{}^{\rm max})$ & $\log({\rm SFR}^{\rm min})$ & $\log({\rm SFR}^{\rm max})$ & $p(M_\ast,{\rm SFR})$\\ 
\cline{1-5}
7.0 & 7.229 & -5.5 & -5.167 & 0.0 \\
7.0 & 7.229 & -5.167 & -4.833 & 0.0 \\
7.0 & 7.229 & -4.833 & -4.5 & 0.0 \\
7.0 & 7.229 & -4.5  & -4.167 & 0.0 \\
7.0 & 7.229 & -4.167  & -3.833 & 0.0 \\
7.0 & 7.229 & -3.833 & -3.5 & 0.0 \\
7.0 & 7.229 & -3.5 & -3.167 & 1.830809731518871e-06 \\
7.0 & 7.229 & -3.167 & -2.833 & 3.876326253400243e-06 \\
7.0 & 7.229 & -2.833 & -2.5 & 6.905726150471546e-06 \\
7.0 & 7.229 & -2.5 & -2.167 & 6.131134402365386e-06 \\
7.0 & 7.229 & -2.167 & -1.833 & 1.555843012065854e-05 \\
7.0 & 7.229 & -1.833 & -1.5 & 1.423655645772669e-05 \\
7.0 & 7.229 & -1.5 & -1.167 & 3.249632877051857e-05 \\
7.0 & 7.229 & -1.167 & -0.833 & 2.3823821312203474e-05 \\
7.0 & 7.229 & -0.833 & -0.5 & 2.2302151013959734e-05 \\
7.0 & 7.229 & -0.5 & -0.167 & 8.486969225985846e-06 \\
7.0 & 7.229 & -0.167 & 0.167 & 0.0 \\
7.0 & 7.229 & 0.167 & 0.5 & 3.829732624985512e-07 \\
7.0 & 7.229 & 0.5 & 0.833 & 0.0 \\
7.0 & 7.229 & 0.833 & 1.167 & 0.0 \\
7.0 & 7.229 & 1.167 & 1.5 & 0.0 \\
7.0 & 7.229 & 1.5 & 1.833 & 0.0 \\
7.0 & 7.229 & 1.833 & 2.167 & 0.0 \\
7.0 & 7.229 & 2.167 & 2.5 &  0.0 \\
\cline{1-5}
\end{tabular}
\end{table*}

\end{document}